\documentclass[prl,a4paper,superscriptaddress,twocolumn]{revtex4}
\usepackage{graphicx,amsmath,bm, color}
\usepackage{latexsym}
\usepackage{amssymb}
\usepackage{amsfonts}
\usepackage{amsthm}
\usepackage{mathrsfs}
\usepackage{natbib}
\usepackage{hyperref}
\usepackage{cleveref}
\usepackage{verbatim,graphics, float}
\usepackage{psfrag}
\usepackage{subfigure}

\newcommand{\ba}{\begin{eqnarray}}
\newcommand{\ea}{\end{eqnarray}}
\newcommand{\ban}{\begin{eqnarray*}}
\newcommand{\ean}{\end{eqnarray*}}
\newcommand{\moy}[1]{\langle #1 \rangle}
\newcommand{\ket}[1]{\mbox{$ | #1 \rangle $}}
\newcommand{\bra}[1]{\mbox{$ \langle #1 | $}}

\begin{document}

\title{Device-independent certification of the teleportation of a qubit}
\date{\today}

\author{Melvyn Ho}
\affiliation{Centre for Quantum Technologies, National University of Singapore, 3 Science Drive 2, Singapore 117543}
\author{Jean-Daniel Bancal}
\affiliation{Centre for Quantum Technologies, National University of Singapore, 3 Science Drive 2, Singapore 117543}
\author{Valerio Scarani}
\affiliation{Centre for Quantum Technologies, National University of Singapore, 3 Science Drive 2, Singapore 117543}
\affiliation{Department of Physics, National University of Singapore, 2 Science Drive 3, Singapore 117542}

\begin{abstract}
We want to certify in a black box scenario that two parties simulating the teleportation of a qubit are really using quantum resources. If active compensation is part of the simulation, perfect teleportation can be faked with purely classical means. If active compensation is not implemented, a classical simulation is necessarily imperfect: in this case, we provide bounds for certification of quantumness using only the observed statistics. The usual figure of merit, namely the average fidelity of teleportation, turns out to be too much of a coarse-graining of the available statistical information in the case of a black-box assessment.

\end{abstract}
\maketitle

\textit{Introduction.--} Shortly after the milestone paper that introduced quantum teleportation \cite{Teleportation93}, the question was asked of which deviation from the ideal case one can tolerate while still claiming that proper quantum effects are being observed. Popescu proved that, if Alice and Bob share no entanglement, the average fidelity for the teleportation of unknown qubit states is bounded by $\bar{F}=\frac{2}{3}$ \cite{Popescu94}. This bound has been widely used as a benchmark for experiments \cite{Bouwmeester97,marcikic03}, including the most recent ones \cite{yin2012,ma2012a}. The bound at $\frac{2}{3}$ holds if one trusts that the quantum systems under study are qubits. By using larger-dimensional alphabets, the model of Gisin \cite{Gisin96} reaches up to an average fidelity $\bar{F}\approx 0.87$ with only classical resources. The relation between Bell's theorem and teleportation, which inspired already \cite{Popescu94}, has been further studied by the Horodecki family \cite{HorodeckiTele}, Zukowski \cite{Zukowski}, and eventually by Clifton and Pope \cite{CliftonPope}. This work states that $\bar{F}\gtrsim 0.9$ would guarantee that the observed teleportation phenomenon has not been simulated with local variables. 

However, at a more careful glance, even the treatments based on Bell inequalities invoke two-qubit algebra at one stage or another to complete the calculation. The fact that it could be critical to resort to qubits at any stage was noticed only several years later, in the context of quantum key distribution \cite{agm06}, where this threatened the security of existing protocols. This observation in turn lead to the idea of \textit{device-independent assessment} \cite{a07}. The device-independent framework has since been applied to several quantum information tasks (see \cite{rmp,slovaca} for reviews). It is time to reconsider the assessment of quantum teleportation in this by now well established framework: this is the goal of the present paper.

\begin{figure}
\includegraphics[width = 8.5cm]{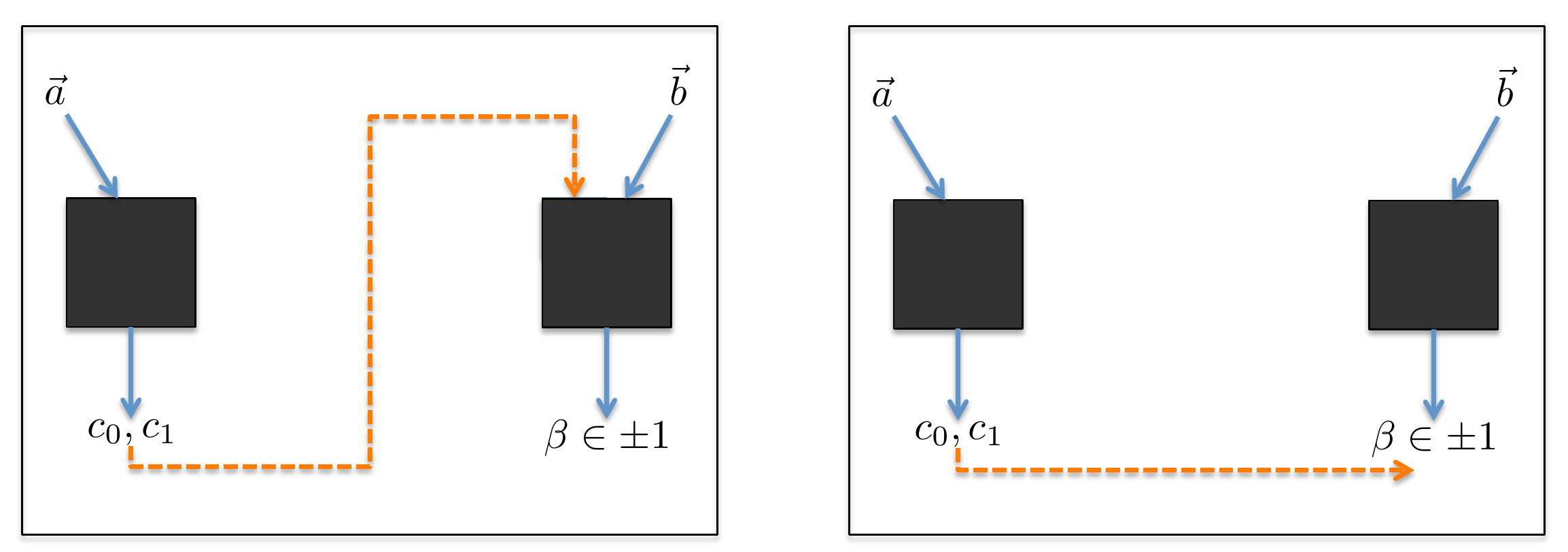}
\caption{Operational black-box description of the teleportation of a qubit in 2 different ways Bob can use the two bits of communication. In the first scenario, the two bits are input into Bob's side to perform a compensation depending on $(c_{0}, c_{1} )$. In the second scenario, Bob's box cannot benefit from the two bits of information, since they are sent only after the outcomes are obtained.   }
\label{fig1}
\end{figure}

\textit{Operational black-box description.--} A vendor is selling two boxes which allegedly perform quantum teleportation of a qubit state (Fig.~\ref{fig1}). The input of each box consists of a unit vector on the surface of sphere $\mathbb{S}^2$ embedded in $\mathbb{R}^3$. Alice's vector $\vec{a}$ is meant to describe the state to be teleported as $\ket{\psi}\bra{\psi}=\frac{1}{2}(\mathbb{I}+\vec{a}\cdot\vec{\sigma})$ \footnote{Teleportation should work also for mixed states, i.e. $|\vec{a}|\leq 1$; since the most demanding simulation is that of a pure state, in this paper we focus on $|\vec{a}|=1$.}. Bob's vector is meant as his choice of performing the measurement along $\vec{b}\cdot\vec{\sigma}$ on the teleported state. Alice and Bob inform the vendor that they will treat their inputs as defined in the same reference frame. For every input, Alice's box outputs two bits $(c_0,c_1)\in\{0,1\}^2$, while Bob's box outputs a bit $\beta\in\{-1,+1\}$. The two boxes are claimed to be loaded in each run with a maximally entangled two-qubit state. But there is no direct \textit{a priori} evidence for it: in fact, this is we will aim to infer \textit{a posteriori} by observing the statistical behavior of the boxes. In the ideal teleportation experiment, conditioned on $(c_0,c_1)$, Bob's box contains a qubit in the state $\rho_B=\frac{1}{2}(\mathbb{I}+(\mathbf{R}_{c_0c_1}\vec{a})\cdot\vec{\sigma})$, where the four $SO(3)$ matrices are the identity $\mathbf{R}_{00}=I$ and the rotations by $\pi$ along three orthogonal directions, $\mathbf{R}_{01}=\mathbf{R}(\hat{x},\pi)$, $\mathbf{R}_{10}=\mathbf{R}(\hat{y},\pi)$ and $\mathbf{R}_{11}=\mathbf{R}(\hat{z},\pi)$.

We are going to show that a black-box certification of qubit teleportation is indeed possible and provide explicit bounds for its conclusiveness. As a first step, we have to spell out two consequences of working in a black-box scenario.

\textit{Consequence 1: impossibility of active compensation.--} When teleportation is used as a building block in a larger protocol, one typically wants to recover the input state of Alice deterministically on Bob's side. To this effect, Alice is asked to send $(c_0,c_1)$ over to Bob's location, so that he can apply the unitary transformation corresponding to $\mathbf{R}_{c_0c_1}$ on the Bloch vector, and ideally recover Alice's state. Experiments that include this active compensation \cite{ma2012a} are rightly considered as more advanced than those that don't. If Bob were to perform a measurement along $\vec{b}$ after the compensation, he expects to find $\moy{\beta}=\vec{a}\cdot\vec{b}$. Now, in the black-box scenario, active compensation translates to allowing two more bits of input in Bob's box (Fig. \ref{fig1}a), but the way this information is processed within the box may be very different from applying a rotation. Thus, the black-box version of teleportation with active compensation gives more leeway for the vendor to cheat --- too much leeway, as it turns out: with those additional two bits of input in Bob's box, the statistics of perfect teleportation can be simulated with only classical resources. This conclusion is an immediate corollary of the Toner-Bacon simulation of the singlet \cite{TonerBacon03}; it supersedes previous, slightly less efficient simulations of teleportation \cite{CGM00}. As we are going to see, black-box certification of teleportation becomes possible if one collects $(c_0,c_1)$ and $\beta$ separately, then studies the conditional statistics: in other words, if one presents the data of teleportation in analogy with a Bell test \cite{Zukowski}.

\textit{Consequence 2: need to assume that the state to be teleported is known.--} One of the main features of quantum teleportation is that the protocol works even when the state to be teleported is unknown to the person who sets up the Bell-state measurement. Whether this feature must be requested of simulations has been debated, and ultimately depends on each author's choice of assumptions. For the black-box scenario, the case is clear: the certification must be done against protocols that simulate teleportation of a \textit{known} state. Indeed, in the two-box scenario presented here, Alice's box simulates both the qubit source and the Bell state measurement. Since Alice inputs $\vec{a}$ as classical information, this information can be copied and made available at any of the internal steps that happen in the box. It would not help to ask the vendor to build the source as a separate box, because this box must send a signal (the alleged qubit) to the box that allegedly performs the Bell state measurement: that signal could carry the classical description of the state.

\textit{Operational description of teleportation.--} After these considerations, we focus on the scenario sketched in Fig.~\ref{fig1}b. The available data are a table of values $(c_0,c_1,\beta|\vec{a},\vec{b})$ for each run of the experiment. For each choice of $\vec{a},\vec{b}\in\mathbb{S}^2$, we assume that infinitely many runs of the experiment are performed (that is, we neglect statistical fluctuations). From this table, one can extract the probability distributions $P(c_0,c_1,\beta|\vec{a},\vec{b})$, which will be used to check the violation of a Bell inequality. The ideal case is represented by
\ba P_{\mathrm{ideal}}(c_0,c_1,\beta|\vec{a},\vec{b})&=&\frac{1}{8}\,[1+ \beta  (\mathbf{R}_{c_o c_1}\vec{a}) \cdot \vec{b}]\,.
\ea The \textit{average fidelity of teleportation} is the most frequently used measure of quality of a teleportation protocol. In order to define it in this operational scenario, notice first that $(\mathbf{R}_{c_o c_1}\vec{V}) \cdot \vec{b} = \vec{V} \cdot (\mathbf{R}_{c_o c_1}\vec{b})$. Thus, Bob will sort the data of the table in order to reconstruct
\ba
P(\beta|c_0,c_1,\vec{a},\mathbf{R}_{c_o c_1}\vec{b})&=&\frac{1}{2}(1+\vec{V'}_{c_o c_1}(\vec{a})\cdot\vec{b})\,.
\ea
On the right-hand side, we have made the assumption (which can be verified a posteriori) that the observed $\moy{\beta}$ is linear in $\vec{b}$. If this were not the case, it is manifest that Bob is not measuring a qubit. If the linear behavior is indeed observed, Bob can extract a compensated vector $\vec{A}_{c_o c_1}(\vec{a})$ which should ideally be equal to $\vec{a}$. The average teleportation fidelity can thus be estimated by sampling Alice's inputs at random:
\ba
\bar{F}&=&\int\frac{d\vec{a}}{4\pi}\sum_{c_0,c_1}P(c_0,c_1)\,\frac{1+\vec{A}_{c_o c_1}(\vec{a})\cdot\vec{a}}{2}\,.
\ea

\textit{Device-independent certification of quantum resources in teleportation.--} Let us move to the constructive description of the certification, based on the scenario .  Here we present one approach, not claimed to be optimal, that uses, like \cite{CliftonPope} , the CHSH inequality. In the protocol, Alice can choose between two inputs $\vec{a}_0$ and $\vec{a}_1$; similarly, Bob can choose between $\vec{b}_0$ and $\vec{b}_1$. Alice's outcome consists of two bits, from which we want to extract one bit $\alpha\in\{-1,+1\}$: we choose the prescription $\alpha\equiv   2c_j -1$ if Alice's input was $\vec{a}_j$.  Now one can evaluate
\ba
\mathrm{CHSH}&=&E_{00}+E_{01}+E_{10}-E_{11}
\ea
with
\ban
E_{jk} &\equiv& P(\alpha = \beta|j,k) - P(\alpha \neq \beta|j,k)\nonumber \\
&=& P(c_j = 0,  \beta= -1|j,k)   + P(c_j = 1,  \beta= +1|j,k) \nonumber \\
&-& P(c_j = 0,  \beta= +1|j,k)-P(c_j = 1,  \beta= -1|j,k) \ean
where $P(c_j,\beta|j,k)\equiv P(c_j,\beta|\vec{a}_j,\vec{b}_k)$. If $\mathrm{CHSH}>2$ in a loophole-free assessment, the two boxes must have shared quantum entanglement. This is a very standard device-independent argument by now. The interesting step consists in studying its implications for a teleportation setup.

\textit{A first application.--} For a first assessment, let us consider a pair of boxes that produces the statistics
\ba P_{\mathrm{obs}}(c_0,c_1,\beta|\vec{a},\vec{b})&=&\frac{1}{8}\,[1+ \beta   \vec{V}_{c_o c_1}(\vec{a}) \cdot \vec{b}]\,.\label{pobslin}
\ea This model captures in particular Alice's statistics $P(c_{0}, c_{1} | \vec{a}) = \frac{1}{4}$, as well as the fact that Bob's statistics are linear in $\vec{b}$ (if this were not the case, Alice and Bob would immediately be suspicious because of a non-trivial departure from the qubit behavior). Then \ba
\mathrm{CHSH} &=& \frac{1}{4} \sum_{c_1}(\vec{b}_0 + \vec{b}_1) \cdot \left[  \vec{V}_{0, c_1}(\vec{a}_0)  - \vec{V}_{1, c_1}(\vec{a}_0) \right] \nonumber \\
&& +  \frac{1}{4} \sum_{c_0}(\vec{b}_0 - \vec{b}_1) \cdot \left[  \vec{V}_{c_0, 0}(\vec{a}_1)  - \vec{V}_{c_0, 1}(\vec{a}_1) \right] \label{GeneralCHSH}
\ea Assume further that
\ba\vec{V}_{c_0, c_1}(\vec{a})=\lambda\mathbf{R}_{c_0c_1}\vec{a}&,&\lambda\in[0,1]\label{stretch}\ea that is, in an active-compensation teleportation Bob would always retrieve $\lambda\vec{a}$ if Alice has input $\vec{a}$, independent of $c_0$ and $c_1$; again, this is the expected behavior when the resource state is a Werner state, for example,  and can be checked on the observed statistics. This assumption also implies that the post-processing fidelity of each vector is the same as the active compensation case. Then
\ba 
\mathrm{CHSH} = \lambda\,\left[ \sum_{c_1}(\vec{b}_0 + \vec{b}_1) \cdot  \vec{a}_{0, x}   +   \sum_{c_0}(\vec{b}_0 - \vec{b}_1) \cdot   \vec{a}_{1, y}  \right]
\label{CHSH} 
\ea
where $\vec{a}_{0, x}$ refers to the x-component of the vector  $\vec{a}_0$, and $ \vec{a}_{1, y}$ refers to the y-component of the vector $\vec{a}_1$. Here, the vectors $\vec{V}_{c_0c_1}$ represent the vector that Bob would have had, without compensation to obtain $\vec{A}_{c_0c_1}$. The fact that the spatial symmetry is broken is just a consequence of our initial choice of $\alpha$, as different choices could be used to pick up any of the following pairs of components from Alice's vector: ($a_x, a_y$), ($a_x, a_z$), ($-a_x, a_y$) and so on. Therefore, if Alice were to choose $\vec{a}_0 = \vec{x}$ and $\vec{a}_1 = \vec{y}$, there are settings for Bob such that the resulting CHSH expression is violated whenever $\lambda \geq \frac{1}{\sqrt{2}}$. This translates to a critical average teleportation fidelity $\bar{F} = \frac{1}{2} (1 + \lambda) \geq \frac{1}{2} (1 + \frac{1}{\sqrt{2}})\approx 0.85$ \footnote{This is precisely the fidelity bound that guarantees nonlocality in the Werner state with the qubit assumption \cite{CliftonPope}. }.

From our calculation, one may be tempted to draw the conclusion that $\bar{F}\geq 0.85$ is sufficient to certify quantum teleportation in a black-box scenario. However, there is a counter-example to this statement: in 1996, Gisin has presented the simulation of the teleportation of a known state that achieves $\bar{F}\approx 0.87$ using only classical resources \cite{Gisin96}. We are going to study this model in the next section.

\textit{Gisin's classical simulation revisited.--} In a classical simulation of teleportation, only the two bits of communication will convey information about $\vec{a}$ to Bob's box. Gisin's protocol uses them to tell in which quarter of the Bloch sphere $\vec{a}$ lies. The Bloch sphere is divided in four equal quarters $S_{i j }$, each having in its center one of the $\vec{t}_{i j}=\mathbf{R}_{ij}\vec{t}_{00}$ thus defined: $\vec{t}_{00} = \frac{1}{\sqrt{3}}(+1, +1, +1)$, $\vec{t}_{01} =\frac{1}{\sqrt{3}} (+1, -1, -1)$, $\vec{t}_{10} = \frac{1}{\sqrt{3}}(-1, +1,-1)$ and $\vec{t}_{11} = \frac{1}{\sqrt{3}}(-1, -1, +1)$. Bob's box contains $\vec{t}_{00}$ and outputs $\beta$ distributed according to $\moy{\beta}=\vec{b}\cdot\vec{t}_{00}$. Thus the protocol produces
\ba P_{\mathrm{Gisin}}(c_0,c_1,\beta|\vec{a},\vec{b})&=&\delta(\vec{a}\in S_{c_0c_1})\,\frac{1}{2}\,(1+ \beta   \vec{t}_{00}\cdot \vec{b})\,.\label{pobsgisin}
\ea Notice how Alice and Bob are completely uncorrelated. These statistics lead to $\textrm{CHSH}=0$. Also, it is easy to see that
\ba P_{\mathrm{Gisin}}(\beta|c_0,c_1,\vec{a},\mathbf{R}_{c_0c_1}\vec{b})&=&\frac{1}{2}\,(1+ \beta \vec{t}_{c_0c_1}\cdot \vec{b})\label{fidgisin}
\ea for $\vec{t}_{c_0c_1}$ which is in the same sector as $\vec{a}$. The average teleportation fidelity is therefore
\ba
\bar{F}  &=& \frac{1}{4}\sum_{c_0 c_1} \frac{ d\vec{a}}{\pi} \int_{S_{c_0c_1}}  \frac{1 + \vec{t}_{c_0 c_1}  \cdot \vec{a}}{2} \,=\, \int_{S_{00}}  \,\frac{ d\vec{a}}{\pi}\,\frac{1 + \vec{t}_{00}  \cdot \vec{a}}{2}  \nonumber \\
 &= &\frac{1}{2} \left[ 1+ 3 \frac{1}{ \pi} \int_{\pi/3} ^{\pi}d\phi \int_{0} ^{u(\phi)}d\theta \cos \theta \sin \theta \right]\,\approx\, 0.87  \nonumber    \label{GisinFidelity}
\ea with $u\equiv \tan^{-1}[\frac{\sqrt{2}}{\cos (\phi+\frac{\pi}{3})}]$. 

The observable statistics \eqref{pobsgisin} are however significantly different from the ones we posited before [Eqs \eqref{pobslin} and \eqref{stretch}]. Notably, Alice's output $(c_0,c_1)$ is deterministic for a given $\vec{a}$. Bob's box is found to contain always the same vector $\vec{t}_{00}$, which does not depend on $\vec{a}$ at all. One could try to modify the protocol in order to erase these obvious shortcomings. For instance, Alice's output could be randomized by adding two bits of shared randomness $(r_0,r_1)$ to both her and Bob's box. Alice's box would then output $(c_0,c_1)=(c_0'\oplus r_0,c_1'\oplus r_1)$ when $\vec{a}\in S_{c_0'c_1'}$, while Bob's box would contain $\vec{t}_{r_0r_1}$. Since $\mathbf{R}_{ij}\mathbf{R}_{i'j'}=\mathbf{R}_{i\oplus i',j\oplus j'}$, this hashing leaves unchanged \eqref{fidgisin}, thence the fidelity. As for \eqref{pobsgisin}, it is replaced by $P_{\mathrm{Gisin}}(c_0,c_1,\beta|\vec{a},\vec{b})=\frac{1}{8}$, and gives CHSH = 0. So at least Alice does not detect anything obviously wrong locally, since $P(c_0,c_1)=\frac{1}{4}$.

One could also randomise Bob's vectors by randomising the frame uniformly of the tetrahedron for each run of the protocol, and for Bob's box to contain $t_{00}$ of the that frame for every run. This also gives us $P(c_0c_1|\vec{a})= \frac{1}{4}$, and nonzero $\vec{V}_{c_0c_1}$ such that one obtains a nontrivial CHSH value, but also a low fidelity of 0.5.

The point here is that in excluding local protocols, the local statistics may already cast doubt on whether the protocol is truly close to the ideal case. Furthermore, the local statistics may be used as verifiable assumptions, to form useful bounds specific for a particular experiment.
For us, the Gisin model and the variants we just discussed serve to highlight the fact that these assumptions should be verified when concluding if the teleportation protocol utilises quantum resources, especially in the fidelity region close to 85\%. Also, despite the high fidelity of the Gisin model in the active case, these simple modifications do not show a strong relation between the post-processing fidelity and CHSH value.

\textit{Low fidelity, high CHSH.--} Now we consider whether it is possible to observe a low teleportation fidelity, and yet a high CHSH violation. 

For this situation, consider a teleportation protocol that maps $\vec{a}$ to the $V_{c_0 c_1}(\vec{a})$ that one expects in a perfect teleportation experiment, but only for the two vectors we require for our CHSH function:

\[
\vec{a} \mapsto \vec{V}_{c_0c_1}(\vec{a})  = 
\begin{cases} 
R_{c_0 c_1}\vec{a}  &  \text{for } \vec{a} \in \{ \vec{x}, \vec{y} \} \\ 
0 & \text{otherwise} 
\end{cases}
\] 

In this case, the average fidelity across the entire sphere is essentially 0.5. However, one can obtain CHSH $= 2 \sqrt{2}$ by using the settings that were previously chosen. For sure there is nonlocality in the system, and this is reflected in the violation of CHSH, but this maximal violation is attainable by a system that has a very bad fidelity, indicating that a low average fidelity might have little relation to the performance of the protocol with respect to a finite number of input choices, and does not necessarily mean that the protocol is local.

While it is not clear that something close to this extreme case can happen in practice, at least the tools we use here do not allow us to put a tighter bound on the lowest fidelity for which a Bell violation can be observed.

\textit{Highest fidelity without CHSH violation.--} To complete our study, we also describe a possible protocol $P_{crit}$ that has the highest average fidelity without yielding a violation. Here we will not impose the condition in Eqn (\ref{stretch}), but only that the compensated $\vec{V}_{c_0 c_1}$ form a consistent description, i.e. $ \forall \, \vec{a_x}  , \exists \,  \vec{a}_x^B $ s.t. $R_{c_0 c_1}  \vec{V}_{c_0 c_1} (\vec{a_x} ) = \vec{a}_x^B $; along with Eqn(\ref{pobslin}). To construct such a protocol, we recall that the CHSH function we derived earlier (in particular, the specific coarse graining to determine $\alpha$) picks out a particular component of each of the teleported $\vec{a}_x^B$. This means that to limit  CHSH $\leq 2$ for all choices of  settings, the resultant vectors $\vec{a}_x^B$ must have individual components limited to some amount. For example, if the  teleported vector $\vec{a}_{0}^B$ is such that  $\vec{a}_{0, z}^B = \pm \frac{1}{\sqrt{2}}$, then $\vec{a}_{1,x}^B$ and $\vec{a}_{1,y}^B$ for any input $\vec{a}_{1}$ must be limited to $\frac{1}{\sqrt{2}}$ as well. We could also limit the maximal z-component of any teleported vector to be an arbitrary $W_z$, meaning that the x and y components of any other vector should be at most $\sqrt{2} - W_z$ to obtain CHSH=2. However, we checked that the protocol with the highest overall fidelity is for $W_z = \frac{1}{\sqrt{2}}$. 

There are two distinct classes of optimal assignments $\vec{a}_x^B$ for every $\vec{a}$. Some vectors could be teleported with perfect fidelity without having any component larger than $\frac{1}{\sqrt{2}}$. One such example of Alice's input is the vector $\vec{a} =  \frac{1}{\sqrt{3}}(+1, +1, +1)$. For such inputs, we can allow the protocol to teleport these vectors perfectly, as such vectors do not yield a violation. Therefore, for inputs of Alice in such a region, we allow our protocol to teleport with perfect fidelity:

\[
 \vec{a} \mapsto  \vec{a}_x^B  = \vec{a} 
\] 

The second class of assignment $\vec{a}_x^B$ is for vectors that would allow for a violation if they were perfectly teleported. To avoid this, we deterministically assign $\vec{a}_x^B$ in such a way as to maximise the fidelity while keeping the largest component at $\pm \frac{1}{\sqrt{2}}$. As an example, consider inputs in the upper cap of the Bloch sphere, with z-component larger than $\frac{1}{\sqrt{2}}$. With such inputs, the teleported vector with the highest fidelity without violation is a vector on the intersection of $z = \frac{1}{\sqrt{2}}$. Hence, our protocol should not be allowed to teleport these vectors perfectly, but with some reduced fidelity:

\[
{\text{Upper cap}}: \vec{a}  =  \left(
 \begin{array}{cc}
\sin \theta \, \cos \phi \\
\sin \theta  \, \sin \phi\\
\cos \theta \\
\end{array} 
 \right)
\mapsto  \vec{a}_x^{B}   = 
 \left(
 \begin{array}{cc}
\frac{1}{\sqrt{2}}  \, \cos \phi \\
\frac{1}{\sqrt{2}} \,  \sin \phi \\
\frac{1}{\sqrt{2}} \\
\end{array} 
 \right)
\]

\begin{figure}[htbp]
\includegraphics[width = 5.0cm]{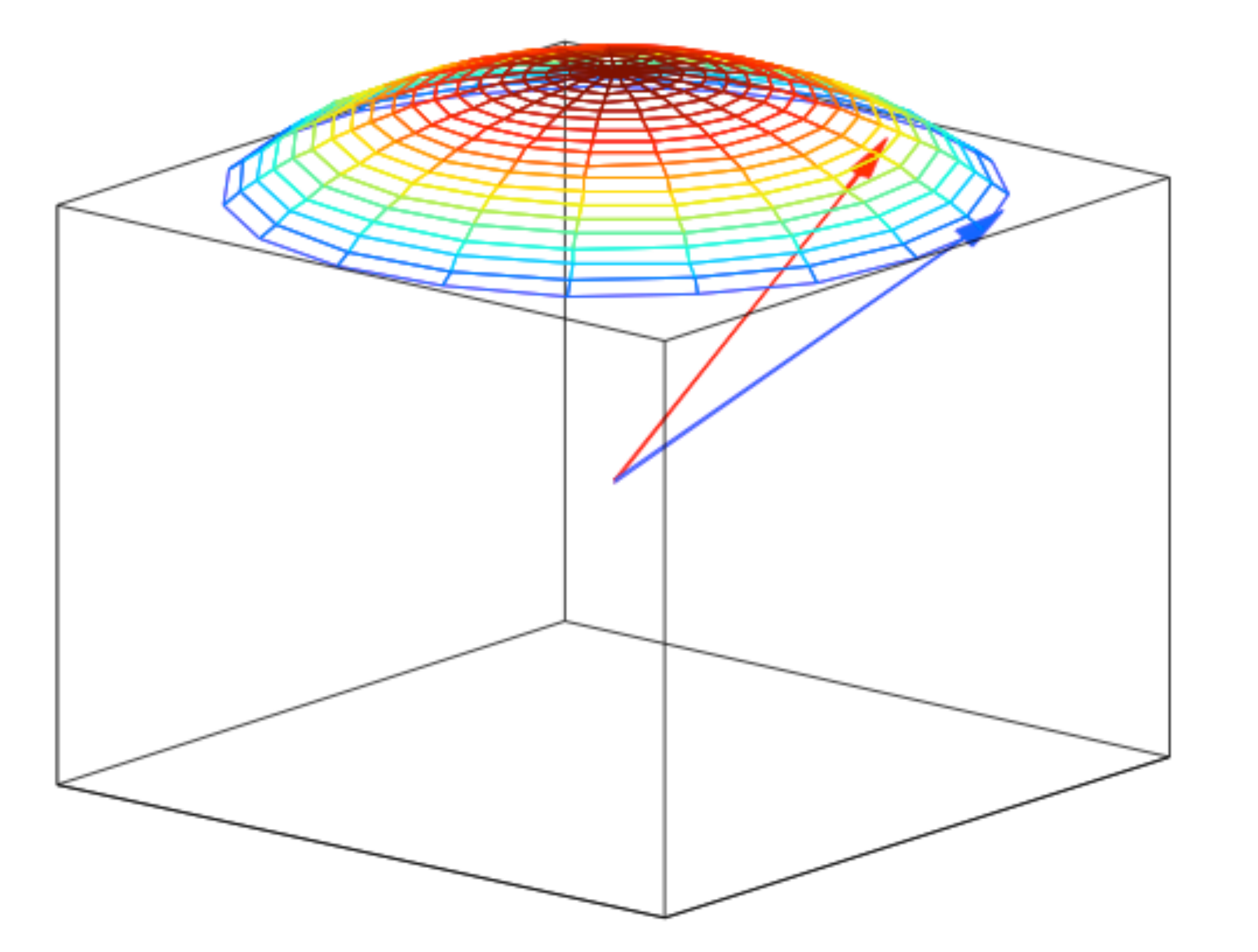}
\caption{Schematic of the region of the sphere with z-component larger than $\frac{1}{\sqrt{2}}$. For a particular input vector (in red), the optimal output vector with the highest fidelity (in blue) would be a vector with the same phase $\phi$, but with z-component limited to prevent any possible violation. }
\label{fig1}
\end{figure}

The other 5 regions are mapped in a similar fashion.

Average fidelity over inputs with z-component larger than $\frac{1}{\sqrt{2}}$:
\ba
\bar{F}_{cap} &=& \frac{1}{2} + \frac{1}{2} \frac{\int_{\phi = 0} ^{2 \pi}   \int_{\theta = 0} ^{\pi/4} \, \vec{a}_x^{B} \cdot  \vec{a} \, \sin \theta \, d \phi \, d \theta}{\int_{\phi = 0} ^{2 \pi}   \int_{\theta = 0} ^{\pi/4} \,  \sin \theta \,  d \phi  \, d \theta } \\
&=& \frac{1}{2} + \frac{1}{2} \frac{        \frac{\pi^2}{4 \sqrt{2}}}{2\pi{(1-\frac{1}{\sqrt{2}})}} \\
&=& \frac{1}{2} \left[1+ \frac{\pi}{8(\sqrt{2} - 1)} \right]
\ea

Assigning a fidelity of 1 to the remaining regions, the fidelity of the protocol $P_{crit}$ is then 
 
 \ba
\bar{F}_{P_{crit}} &=&  \frac{ [(12 \pi) (1-  \frac{1}{ \sqrt{2}})]   *\bar{F}_{cap}       +  [4 \pi - ( 12 \pi) (1-  \frac{1}{ \sqrt{2}})  ] }{4 \pi}\nonumber \\
&\approx& 0.97718
 \ea

\section{Summary}

In the first section, we investigate and lay out the framework of quantum teleportation in the device-independent scenario, and point out that with active compensation, the average fidelity is not suitable as an indicator of nonlocality. We thus propose the use of the post-processing fidelity, which does not allow the users to benefit from communication in order to producing their outputs. This, with some verifiable assumptions on the local probability distributions, allows us to construct a CHSH-type expression for the outcomes of our teleportation experiment. Here we find that an average fidelity of  85\% and 97.7\% in the post processing scenario is sufficient to quantify nonlocality for different assumptions.  We also explore some other models to see how they perform with respect to our assumptions and the use of the average fidelity. For Gisin's model and its variants, we do not observe any strong relation between fidelity and CHSH. We also give an example with high CHSH and low fidelity to illustrate a possible limitation in using the average fidelity to obtain bounds on the system.

\paragraph{Acknowledgments} 
This work is funded by the Singapore Ministry of Education (partly through the Academic Research Fund Tier 3 MOE2012-T3-1-009) and the Singapore National Research Foundation.\\\\\\

\newpage

\begin{appendix}

\end{appendix}

\end{document}